\documentclass{article}
\usepackage{setspace}
\usepackage{fancyhdr}
\usepackage{booktabs}
\usepackage[margin=1in, headheight=15pt]{geometry}
\usepackage{lipsum}
\usepackage{ragged2e}

\usepackage{graphics}
\usepackage{tikz}
\usetikzlibrary{calc}
\usetikzlibrary{decorations.pathreplacing}
\usetikzlibrary{arrows.meta}
\usepackage{amsmath,amsfonts,amsthm}
\usepackage{mathrsfs}
\usepackage{cancel}
\usepackage[normalem]{ulem}
\usepackage{siunitx}
\usepackage{float}
\usepackage{enumitem}
\usepackage{colortbl}
\usepackage{subcaption} 
\usepackage[font=footnotesize,labelfont=bf]{caption}
\usepackage{tabularx}
\newcolumntype{L}[2]{>{\hsize=#1\hsize\columncolor{#2}\raggedright\arraybackslash}X}%
\newcolumntype{R}[2]{>{\hsize=#1\hsize\columncolor{#2}\raggedleft\arraybackslash}X}%
\newcolumntype{C}[2]{>{\hsize=#1\hsize\columncolor{#2}\centering\arraybackslash}X}%

\usepackage{color}
\definecolor{mygreen}{rgb}{0,0.6,0}
\definecolor{mygray}{rgb}{0.5,0.5,0.5}
\definecolor{mymauve}{rgb}{0.58,0,0.82}
\definecolor{mylightgray}{gray}{.9}
\definecolor{myblue}{rgb}{.28,.24,.55}
\definecolor{mylightblue}{rgb}{0.74, 0.83, 0.9}
\definecolor{mybrightred}{rgb}{1,.13,.32}
\definecolor{mypink}{rgb}{0.96, 0.76, 0.76}
\definecolor{mylightgreen}{rgb}{0.66, 0.89, 0.63}
\usepackage{colorspace}
\definespotcolor{clear}{rgb}{0,0,0,0}

\usepackage{natbib}
\setcitestyle{aysep={}}

\PassOptionsToPackage{hyphens}{url}
\usepackage{hyperref}
\hypersetup{colorlinks=true, citecolor=darkgray, linkcolor=darkgray, urlcolor=darkgray}
\makeatletter
\let\@fnsymbol\@arabic
\makeatother

\usepackage{lmodern}
\usepackage{listings}
\usepackage[textsize=tiny, backgroundcolor=mypink, linecolor=red]{todonotes}

\newcommand{\mytitle}{\textbf{The Russian invasion of Ukraine selectively depolarized the Finnish NATO discussion}}
\newcommand{\shorttitle}{The Russian invasion of Ukraine selectively depolarized the Finnish NATO discussion}
\author{Yan Xia$^{1}$,
Antti Gronow$^{2}$,
Arttu Malkamäki$^{2}$,
Tuomas Ylä-Anttila$^{2}$, \\
Barbara Keller$^{1}$,
and Mikko Kivel\"a$^{1}$ \\\\
\small $^{1}$Department of Computer Science, Aalto University, Finland \\
\small $^{2}$Faculty of Social Sciences, University of Helsinki, Finland \\
}
\newcommand{\surname}{Xia \textit{et al}.}

\title{\mytitle}
\date{}

\begin{document}
\pagenumbering{roman}
\singlespacing
\maketitle
\thispagestyle{empty}
\begin{abstract}
\noindent The Russian invasion of Ukraine in 2022 dramatically reshaped the European security landscape. In Finland, public opinion on NATO had long been polarized along the left-right partisan axis, but the invasion led to a rapid convergence of the opinion toward joining NATO. We investigate whether and how this depolarization took place among polarized actors on Finnish Twitter. By analyzing retweeting patterns, we find three separated user groups before the invasion: a pro-NATO, a left-wing anti-NATO, and a conspiracy-charged anti-NATO group. After the invasion, the left-wing anti-NATO group members broke out of their retweeting bubble and connected with the pro-NATO group despite their difference in partisanship, while the conspiracy-charged anti-NATO group mostly remained a separate cluster. Our content analysis reveals that the left-wing anti-NATO group and the pro-NATO group were bridged by a shared condemnation of Russia's actions and shared democratic norms, while the other anti-NATO group, mainly built around conspiracy theories and disinformation, consistently demonstrated a clear anti-NATO attitude. We show that an external threat can bridge partisan divides in issues linked to the threat, but bubbles upheld by conspiracy theories and disinformation may persist even under dramatic external threats.
\end{abstract}

\pagenumbering{arabic}
\setcounter{page}{1}

\section*{Introduction}
\label{sec:intro}
Despite a period of momentum building, the Russian invasion of Ukraine on Feb 24, 2022 came as a shock to most observers. The shock was most acute in Ukraine but was indirectly felt also in countries bordering Russia. Finland, the militarily non-aligned European country that shares a 1344-kilometer border with Russia, witnessed a sharp shift in its public opinion on NATO membership, based on a reappraisal of the external threat posed by Russia. Traditionally, around 20–30 percent of the Finnish population have been in favor of joining NATO. After the invasion, support for joining NATO soared into as high as 70–80 percent \citep{yle2022poll}.

Behind this major change in opinion, the rising external threat seems to have had a depolarizing effect on the Finnish NATO discussion. For long, Finnish opinions on NATO embodied a polarization that was largely partisanship-based: voters of the main right-wing party (National Coalition) were largely in favor of joining, whereas voters of left-wing parties were the most vocal opponents of NATO \citep{forsberg2018finland}. After the invasion, however, many left-wing supporters changed their opinion, and eventually the Finnish parliament almost unanimously voted in favor of joining NATO (188 for, 8 against).  

Social media opens an unobtrusive observation window \citep{barbera2020use} into whether and how this depolarization took place among the more politically active and partisan segment of the population \citep{ruoho2019inner,bail2021breaking}, including political elites who often play an important role in steering the discussion \citep{matsubayashi2013politicians,barbera2018new}, as well as fringe communities that subscribe to conspiracy theories and disinformation \citep{ferrara2020characterizing,erokhin2022covid}. The digital traces of user interactions make it possible to measure structural polarization by constructing endorsement networks of individuals and observing cohesive groups in them \citep{conover2011political,barbera2015tweeting,garimella2018quantifying,cossard2020falling}, and provide insight into the information spreading and user interaction dynamics that drive opinion (de)polarization \citep{morales2015measuring,esteve2022political}. While network analysis can reveal the structure of user interactions and how it changes over time \citep{chen2021polarization}, content analysis can uncover how the discussion climate evolves and what arguments connect or distinguish opposing sides \citep{weber2013secular,borge2015content}.

Using a combination of network analysis and content analysis methods, we inspect how the Russian invasion of Ukraine changed the polarization dynamics of the Finnish NATO discussion on Twitter. We find that depolarization took place also among Twitter actors, but in a selective manner. Surprisingly, the external threat largely depolarized partisan actors, who are likely to rely on motivated reasoning and interpret information on the basis of previous opinions \citep{mullinix2016partisanship}; instead, the polarization that persisted was led by actors who built arguments around conspiracy theories and disinformation. Our study therefore adds new empirical evidence to the long-held theory that external conflicts increase internal cohesion \citep{coser1956functions} in how this process may happen selectively, while complementing previous empirical research on the rally-around-the-flag effect \citep{chowanietz2011rallying,steiner2023rallying} and political depolarization \citep{myrick2021external} in the face of external threats.

\section*{Data and Methods}
\label{sec:method}
We collected all tweets in Finnish language from Dec 30, 2021 to Mar 30, 2022 that contain any NATO-related keyword (see Appendix for the list of keywords) using Twitter API v2 \citep{twitter2022twitter}. This gave us 320,407 tweet records in total. We divided the timeline into two-week periods before Feb 24 and one-week periods after Feb 24, in consideration of the asymmetric activity level before and after the Russian invasion of Ukraine. For each period, we constructed a retweet network of users, where a directed link with weight $w$ connects user $A$ to user $B$ if $A$ retweeted $B$ $w$ times within the period. We used only retweet records (not including quote retweets) for constructing the user networks, as retweet is a relatively certain indicator of endorsement-based connection \citep{metaxas2015retweets}. Following prior work on quantifying polarization in Twitter data \citep{garimella2018quantifying, chen2021polarization}, we used only the largest connected component of each network for subsequent analysis. 

Based on an observation of the retweet networks, we decided to focus our analysis mainly on four periods that are representative of the evolving retweeting dynamics: \textit{before} (Feb 10 to Feb 23, 31,399 total tweets, 12,891 retweets), \textit{right-after} (Feb 24 to Mar 2, 81,433 total tweets, 39,936 retweets), \textit{1-week-after} (Mar 3 to Mar 9, 49,585 total tweets, 23,365 retweets), and \textit{4-weeks-after} (Mar 24 to Mar 30, 20,792 total tweets, 9,103 retweets). The retweet network (largest connected component) contains 3,836 users and 10,774 links in the \textit{before} period, 8,986 users and 32,454 links in the \textit{short-after} period, 6,173 users and 19,309 links in the \textit{after} period, and 3,383 users and 7,598 links in the \textit{long-after} period. 

In order to track stance changes induced by the invasion, we performed our analysis on users who were active in the \textit{before} network. Using the Leiden graph partitioning algorithm \citep{traag2019louvain}, we find clusters of users in the \textit{before} network who retweet mainly within their cluster, and inspect how these users change their behavior and stances in the subsequent periods. We ran the Leiden algorithm 50 times and picked the partitioning that gives the shortest description length of the data under the microcanonical stochastic block model \citep{peixoto2017nonparametric}. For each of the four time periods, we then calculated for each user cluster the number of external retweets, the number of internal retweets, and the external-internal (E/I) ratio \citep{krackhardt1988informal}, in order to examine the change in intra-cluster and inter-cluster communication dynamics.

To get a sense of the evolving discussion climate, we sampled a number of tweets from the data for manual content analysis. For each cluster and each of the four time periods, we randomly sampled 42 tweets from those that got retweeted at least once in the cluster in the period, which resulted in 504 sampled tweets (see Appendix for more sampling statistics). We preferentially sampled tweets that were popular within each cluster by setting the sampling probability of each tweet proportional to its number of in-cluster retweets in the period. A group of four coders (all co-authors) then labeled the stance of each tweet to be pro-NATO, anti-NATO, unclear, or unrelated to NATO. The coders were split into two teams, with two coders on each team. From the 504 tweets in total, 24 tweets were randomly sampled for both teams to code; for the remaining 480, one team coded half and the other team coded the remaining half. Within each team, each coder first labeled the 264 tweets independently, then the two coders discussed cases of disagreement and reached a consensus as a team. The inter-team agreement for the 24 double-coded tweets, as evaluated by Krippendorff's alpha \citep{hayes2007answering}, is 0.80. 

\section*{Results}
\label{sec:res}
\begin{figure*}
\centering
\includegraphics[width=\linewidth]{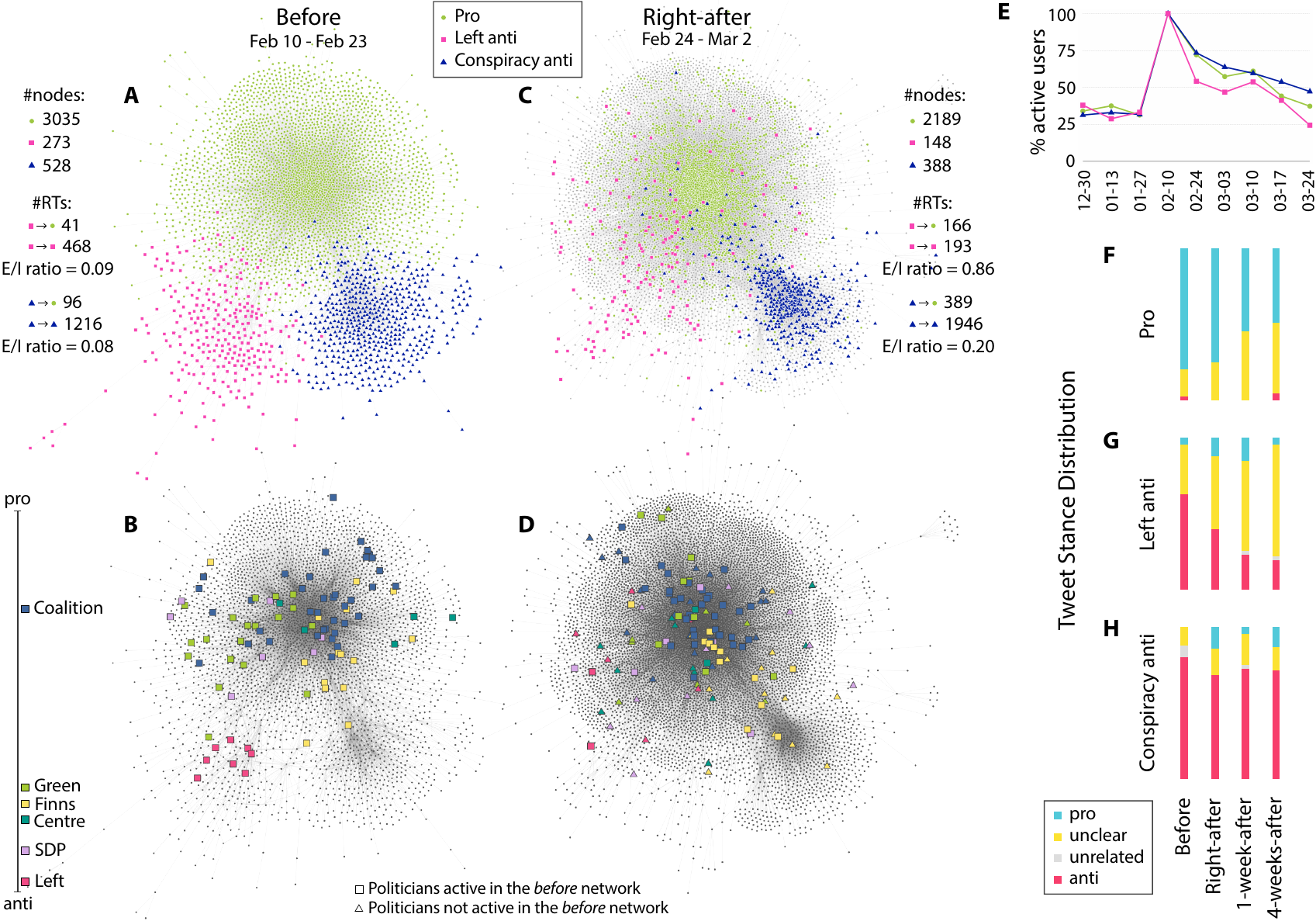}
\caption{Retweet networks and tweet stance distributions showing how the Russian invasion of Ukraine connected the left anti group to the pro group, while the conspiracy anti group persisted. Retweet networks (A)-(B) before and (C)-(D) right after the invasion of Ukraine. Node colors in (A) and (C) correspond to the three groups detected in the \emph{before} network, and the statistics beside each network show the number of nodes, external retweets of the pro group, internal retweets, and the E/I ratio in each anti group. Node colors in (B) and (D) denote the party affiliation of politicians. Parties in the legend are positioned based on the mean of their candidates' attitude toward NATO in 2019, according to an election poll conducted by the Finnish Broadcasting Company (Yle). (E) Change in percentage of active users in each group. Change in distribution of stances on joining NATO among sampled tweets in the (F) pro, (G) left anti, and (H) conspiracy anti group.}
\label{fig:nato}
\end{figure*}

The graph partitioning algorithm reveals three clusters of users in the \textit{before} network (Fig.~\ref{fig:nato}A). Based on the coded stances of sampled tweets in each user group, we find one of the groups to be pro-NATO and the other two to be anti-NATO (Fig.~\ref{fig:nato}F-H). A qualitative reading of the sampled tweets suggests that one of the anti-NATO groups based their arguments on traditional leftists' concerns, such as pacifism and feminism not being compatible with joining a military alliance, and NATO having been involved in violation of human rights. The other anti-NATO group showed a clear engagement in conspiracy theories and disinformation in framing their opposition to NATO. For example, they claimed ``NATO equals supporting `globalism', the global elite, and the World Economic Forum, all of which are supposed co-conspirators that are set out to destroy the Finnish nation'', and that ``those who want people to inject themselves with `poisonous vaccines' are the ones who want to join NATO''.

Our user partisanship analysis further enriches the profile of each group. Specifically, we plot a list of Finnish politician accounts in the \textit{before} network, colored by their publicly available party affiliation (Fig.~\ref{fig:nato}B). In our analysis, we focus on the six main parties in Finland, each with over 10 Members in the current Finnish Parliament: the Left Alliance (Left), the Social Democratic Party (SDP), the Green League (Green), the Centre Party (Centre), the National Coalition Party (Coalition), and the Finns Party (Finns). The parties are ordered by political leaning from left to right. Quite surprisingly, we find that politicians of most parties, including many that traditionally took a neutral or an anti-NATO stance, already fell on the pro-NATO side in the \textit{before} network; presumably, this results from the buildup to the war since the end of 2021. However, politicians affiliated with the Left Alliance -- the traditionally most anti-NATO party -- still fell exclusively in the left anti group. Meanwhile, the conspiracy anti group seems to accommodate few politicians, which suggests its relatively fringe position in political communication.

\subsection*{Change in network structure}
Plotting the pro, left anti, and conspiracy anti users in the retweet networks after the invasion, we observe a significant change in the network structure. In the \textit{right-after} network, members of the left anti group became much less connected internally and more connected to the pro-NATO side, while most members of the conspiracy anti group largely remained in their own internally connected bubble (Fig.~\ref{fig:nato}C). This observation is confirmed by the number of external retweets of the pro group, the number of internal retweets, and the external-internal (E/I) ratio in each anti group: although the E/I ratio of the conspiracy anti group also more than doubled in the first week after the invasion (some of this change might be explained by overfitting \citep{salloum2022separating}), the E/I ratio of the left anti group had an almost tenfold increase in the same period. It is also worth noting that the structural change in the \textit{right-after} retweet network also remains in the \textit{1-week-after} and \textit{4-weeks-after} periods.

The change in retweet network structure reflects a breakage of the cohesive cluster formed by the left anti users, as they instantly developed connection and alignment with the pro group after the invasion. The partisanship plot after the invasion (Fig.~\ref{fig:nato}D) confirms that the invasion bridged the communication divide between politicians of the Left Alliance and those of the other parties. By contrast, the sustained bubble structure of the conspiracy anti group suggests that the invasion did not change its communication dynamics as much.

\subsection*{Change in discussion climate}
Our reading of the sampled tweets suggests that the left anti group shared with the pro group a critical attitude toward Russia's invasion of Ukraine, which potentially connected them in the retweet network. After the invasion, many people in the left anti group also moved away from explicitly voicing anti-NATO stances to asking for more discussion on NATO, in addition to arguing that NATO opponents should not be ostracized. Although this might imply that they did not shift their opinion completely toward the other end, the change in their expression opened up a possibility for their interaction with the pro group, as some NATO supporters also agreed that an open discussion involving both sides is acceptable. Thus, the left anti group and the pro group were also connected by a shared understanding of the democratic norms of discussion.  

Meanwhile, members of the other anti group consistently built explicitly anti-NATO arguments upon conspiracy theories and disinformation. Many were also repeating messages of the official Russian propaganda, and some of them, as well-known figures in the Finnish disinformation and conspiracy theory scene, have been interviewed on Russian state TV as supposed experts. Thus, this conspiracy-charged and pro-Russia group presumably did not find much common ground with the pro group, and was not changed much by the invasion.

The stance distribution of the sampled tweets confirms that the conspiracy anti group held a consistently strong anti-NATO attitude even after the invasion (Fig.~\ref{fig:nato}H). Meanwhile, the left anti group saw a notable decrease in the expression of anti-NATO attitude after the invasion (Fig.~\ref{fig:nato}G), yet it also did not turn clearly pro-NATO (see Appendix for a discussion of tweets with unclear stance). This change potentially reflects some extent of self-censorship in this group: while many users might have retained an anti-NATO leaning, they avoided stating anti-NATO stances explicitly after becoming a minority in the discussion.

Our user activity analysis reveals another possible form of self-censorship in the left anti group. Before the invasion, the two anti-NATO groups had a comparable percentage of active users in each retweet network (Fig.~\ref{fig:nato}E); yet after the invasion, the percentage was consistently lower in the left anti group. The partisanship plot (Fig.~\ref{fig:nato}D) also shows that only two of the eight Left Alliance politicians in the \textit{before} network were still present in the \textit{right-after} network. Coupled with the change in retweeting dynamics, the decreased user activity in the left anti group hints at a spiral of silence \citep{noelle1974spiral} among a part of the left anti users, who may have chosen not to share their opinions in response to the shifted discussion climate. 

\section*{Discussion}
\label{sec:disc}
Our analyses provide an overview of how the Russian invasion of Ukraine selectively depolarized the Finnish NATO discussion on Twitter: the left-wing anti-NATO users broke out of their retweeting bubble and connected with the traditionally right-wing pro-NATO group based on established common ground, but the conspiracy-charged anti-NATO group mostly remained a densely connected cluster of its own and persisted in holding an anti-NATO attitude.

Our study sheds light on how a dramatic external threat can change the discussion dynamics between partisan actors. While previous empirical research has found that terrorist and other direct attacks can lead to a rally-around-the-flag phenomenon where the political elite presents a united front \citep{chowanietz2011rallying}, there is less evidence that an indirect external threat posed by an adversarial state would decrease political polarization between partisan actors \citep{myrick2021external}. Our results show that polarization in partisanship-divided issues can be weakened overnight by a dramatic external threat, as actors of opposite leanings build connections on the basis of a shared target of criticism (Russia) and a shared understanding of democratic norms (discussion about, and even opposition of NATO are part of democracy). 

While the observed depolarization of the endorsement network and the change in expressed stances are conclusive, with observational data it is difficult to gauge the amount of actual opinion change or the level of ideological depolarization at large, especially given the possibility of a spiral of silence. Nevertheless, even when depolarization takes the form of self-censored opposition \citep{brody1989policy}, it can still create opportunities for information exposure and conversation between different bubbles, which can serve as a first step toward actual ideological depolarization \citep{mutz2002cross}. 

Our results also echo the finding of previous research that consumers of conspiracy theories tend to concentrate on within-group content and interaction \citep{bessi2015science,douglas2019understanding}, while further adding that groups formed around conspiracy theories and disinformation may be resistant to changing their opinions even in the face of dramatic external threats and bridged partisan divides. This implies that the polarization led by conspiracy and disinformation consumers might be more entrenched than partisanship-based polarization, so that even external threats may not pave a way toward conversation and consensus with these actors.

\section*{Appendix}
\label{sec:app}
\subsection*{Data collection keywords}
Two of the authors who are experts on Finnish politics developed a list of keywords related to the Finnish NATO discussion: \textit{liittoutua, liittoutumaton, liittoutumattomana, liittoutumattomuuden, liittoutumattomuus, liittoutuminen, liittoutumisen, nato, nato-kumppani, nato-kumppanien, nato-kumppanit, nato-kumppanuus, nato-yhteistyö, nato-yhteistyön, nato-yhteistyössä, nato-yhteistyötä, naton, natoon, natossa, natosta, puolustusliiton, puolustusliitosta, puolustusliitto, puolustusliittoon, sotilasliiton, sotilasliitosta, sotilasliitto, sotilasliittoon, suominatoon, natojäsenyyttä, natojäsenyyden, nato-trolli, nato-trollit, nato-trollien, nato-trollaajat, nato-trollaajien, nato-kiima, nato-kiiman, nato-kiimailijat, nato-kiimailijoiden, natoteatteri, natoteatteria,} and \textit{natoteatterista}. 

\subsection*{Tweet sampling statistics}
For each group and each period, we sampled 42 tweets from those that got retweeted at least once in the group in the period. In total, 1,800/221/416 tweets in the \textit{before} period, 4,188/343/1,118 tweets in the \textit{right-after} period, 2,698/257/779 tweets in the \textit{1-week-after} period, and 1,022/88/481 tweets in the \textit{4-weeks-after} period got retweeted at least once in respectively the pro, left anti, and conspiracy anti group.

\subsection*{Tweets with unclear stance}
In our tweet stance coding, a tweet is labeled ``unclear'' if it does not explicitly express a positive or negative attitude toward NATO. Thus, the label ``unclear'' does not necessarily imply an ambiguous attitude toward NATO, but rather that the tweet does not clearly indicate any attitude. For example, tweets labeled ``unclear'' are often reactions to what was currently taking place in the Ukraine war or in the Finnish NATO policy process.

More specifically in the pro-NATO group, many tweets were labeled ``pro'' in the earlier periods because they were advocating for two citizen initiatives that were pro-NATO; but later on, these initiatives became irrelevant because the needed signatures were collected, and the NATO policy process moved on. Thus in later periods, many clearly pro-NATO tweets disappeared from the pro-NATO group and, for example, many tweets condemning Russia's actions in Ukraine took their place. The latter are often labeled ``unclear'' as they are less clearly in favor of NATO, even though such a stance might be implicit. In general, the increase of tweets with unclear stance does not suggest that the group moved toward an ambiguous stance on NATO.

\section*{Acknowledgements}
We want to thank Ted Hsuan Yun Chen and Risto Kunelius for giving extremely insightful feedback on our study. This work was supported by the Academy of Finland (320780, 320781, 332916, 349366, 352561), the Kone Foundation (201804137), and the Helsingin Sanomat Foundation (20210021).

\small\singlespacing\RaggedRight
\Urlmuskip=0mu plus 1mu\relax
\bibliography{references}

\end{document}